\documentclass[graybox]{svmult}
\usepackage{mathptmx}       
\usepackage{helvet}         
\usepackage{courier}        
\usepackage{makeidx}         
\usepackage{graphicx}        
\usepackage{multicol}        
\usepackage[bottom]{footmisc}
\usepackage{amsmath,amssymb,amsfonts}        
\usepackage{hyperref}
\usepackage{cleveref}
\usepackage{comment}
\usepackage{tikz}
\usetikzlibrary{graphs,patterns,decorations.markings,arrows,matrix}
\allowdisplaybreaks
\makeindex             

\def\e{{\rm e}}
\def\ii{{\rm i}}

\def\be{\begin{equation}}
\def\ee{\end{equation}}
\def\bea{\begin{eqnarray}}
\def\eea{\end{eqnarray}}

\def\x{x_{\rm c}}

\newcommand{\ba}{\begin{array}}
\newcommand{\ea}{\end{array}}
\newcommand{\G}{\Gamma}
\newcommand{\Gd}{\Gamma^{\dag}}

\newcommand{\Qhd}{\hat{Q}^{\dag}}
\newcommand{\Qd}{Q^{\dag}}

\newcommand{\Qtd}{\tilde{Q}^{\dag}}
\newcommand{\cd}{c^{\dag}}

\newcommand{\ket}[1]{\left|#1\right\rangle}

\begin{document}
\title*{A curious mapping between supersymmetric quantum chains}
\author{Gyorgy Z. Feher, Alexandr Garbali, Jan de Gier and Kareljan Schoutens}
\institute{Gyorgy Z. Feher \at BME ``Momentum'' Statistical Field Theory Research Group
1111 Budapest, Budafoki \'ut 8, Hungary, \email{g.feher@eik.bme.hu}
\and Alexandr Garbali \at ARC Centre of Excellence for Mathematical and Statistical Frontiers (ACEMS), School of Mathematics and Statistics, University of Melbourne, Victoria 3010, Australia, \email{alexandr.garbali@unimelb.edu.au}
\and Jan de Gier \at ARC Centre of Excellence for Mathematical and Statistical Frontiers (ACEMS), School of Mathematics and Statistics, University of Melbourne, Victoria 3010, Australia, \email{jdgier@unimelb.edu.au}
\and Kareljan Schoutens \at Institute for Theoretical Physics Amsterdam and Delta Institute for Theoretical Physics
University of Amsterdam, Science Park 904, 1098 XH Amsterdam, The Netherlands,
 \email{c.j.m.schoutens@uva.nl}}
%
%
\maketitle

\abstract{
We present a unitary transformation relating two apparently different supersymmetric lattice models in one dimension. The first \cite{FS07} describes semionic particles on a 1D ladder, with supersymmetry moving particles between the two legs. The second \cite{GFNR15} is a fermionic model with particle-hole symmetry and with supersymmetry creating or annihilating pairs of domain walls. The mapping we display features non-trivial phase factors that generalise the sign factors occurring in the Jordan-Wigner transformation. \\[3mm]
We dedicate this work to our friend and colleague Bernard Nienhuis, on the occasion of his 65-th birthday.}

\section{Introduction}\label{sec:intro}

The concept of supersymmetry was conceived in the realm of (high-energy) particle physics, where it expresses a fundamental symmetry between bosonic and fermionic (elementary) particles or excitations in a quantum field theory or string theory. It would seem that in the context of (low-energy) condensed matter systems a similar concept is out of place as superpartners to, say, the electron, if such exist at all, are far out of sight. Nevertheless, we have learned that supersymmetry can be a useful ingredient in relatively simple model systems describing a condensed phase of matter. As soon as the relevant degrees of freedom are not all bosonic, the notion of a fermionic symmetry becomes feasible. 

A particularly simple supersymmetric lattice model, commonly refered to as the M$_1$ model, was proposed in \cite{FSB03}.  It features itinerant spin-less fermions on a lattice (or graph), with supersymmetry adding or taking out a single fermion. Denoting the supercharges as $Q^\dagger$ and $Q$, the Hamiltonian of what is called ${\cal N}=2$ supersymmetric quantum mechanics \cite{W82} is defined as
\begin{align}
H=\{ Q^\dagger, Q \}.
\end{align}
In the M$_1$ model the non-trivial nature of $H$ is induced by stipulating that fermions are forbidden to occupy nearest neighbour sites on the lattice. These simple definitions lead to surprisingly rich and diverse phenomena. On a 1D lattice, the M$_1$ model was found to be critical, and described by the simplest unitary minimal model of ${\cal N}=2$ superconformal field theory \cite{FSB03}. On 2D lattices, there is the remarkable phenomenon of {\it superfrustration}: an extensive (in the area, that is to say the number of sites) entropy for zero-energy supersymmetric ground states \cite{FS05,E05,HS10}.

Additional features of the M$_1$ model in 1D are integrability by Bethe Ansatz and the existence of a mapping to the $XXZ$ model at anisotopy $\Delta=-1/2$ \cite{FNS03}. These features were generalized to a class of models called M$_k$, where up to $k$ fermions are allowed on consecutive lattice sites \cite{FNS03}. At critical behaviour of these models is captured by the $k$-th minimal model of ${\cal N}=2$ superconformal field theory, while massive deformations give rise to integrable massive ${\cal N}=2$ QFT's with superpotentials taking the form of Chebyshev polynomials \cite{FS17}. 

This paper is concerned with two other, and seemingly different, incarnations of supersymmetry in one spatial dimension. The first is a model, proposed by Fendley and Schoutens (FS) \cite{FS17}, where the supercharges $Q$ and $Q^\dagger$ move particles between two legs of a zig-zag ladder. This would suggest that the particles on the two legs be viewed as bosonic and fermionic, respectively, but the situation in the FS model is different: the phases between the (fermionic) supercharges and the particles are such that the particles on the two legs are naturally viewed as anyons with statistical angle $\pm \pi/2$, that is, as semionic particles. Interestingly, pairs of semions on the two legs can form zero-energy `Cooper pairs' and the model allows multiple supersymmetric groundstates that are entirely made up of such pairs. The FS model is integrable by Bethe Ansatz and has a close-to-free-fermion spectrum: all energies agree with those of free fermions on a single chain, but the degeneracies are different.

The second model we discuss was introduced by Feher, de Gier, Nienhuis and Ruzaczonek (FGNR) \cite{GFNR15}. It can be viewed as a particle-hole symmetric version of the M$_1$ model, where the `exclusion' constraint on the  Hilbert space has been relaxed and where the supercharges are now symmetric between particles and holes. In this model fermion number conservation is violated as terms creating and annihilating pairs of fermions are included in the Hamiltonian. The FGNR can be conveniently described in terms of domain walls between particle and hole segments, as the number of such walls is conserved. Also this model allows a Bethe Ansatz solution and the spectrum of the periodic chain has been shown to have characteristic degeneracies. Just as in the FS model, the degeneracies in the FGNR model can be explained by the formation of zero-energy `Cooper pairs'.

The sole purpose of this contribution to the 2017 MATRIX Annals is to establish a unitary transformation between the FS and FGNR models on an open chain. Based on the similarity of the Bethe ansatz solutions and that of the physical properties the existence of such a map is not too surprising. Nevertheless, the details are quite intricate. This holds in particular for the phase factors involved in the mapping, which achieve the task of transforming domain walls in the FGNR formulation to particles which, in the FS formulation, are best interpreted as semions. The non-local patterns of the phase factors can be compared to the `strings' of minus signs featuring in the Jordan-Wigner transformation from spins to (spin-less) fermions.  

\section{Models}\label{sec:models}
In this section, we define the models \cite{FS07, GFNR15}. We refer to the model of \cite{FS07} as FS model, and the model of \cite{GFNR15} as FGNR model.
Both are spinless fermion models on a chain of length $L$ with some boundary conditions. The fermionic creation and annihilation operators $c_i,\cd_i$ ($i=1,\dots,L$) satisfy the usual anticommutation relations
\begin{equation}
\{c_i^\dagger,c_j\}=\delta_{ij},\qquad \{c_i^\dagger,c_j^\dagger\}=\{c_i,c_j\}=0.
\end{equation}
Based on the fermionic creation operators, the on site fermion-number and hole-number operators are defined as
\begin{equation}
n_i = c_i^\dagger c_i \qquad p_i = 1-n_i.
\end{equation}
These operators act in a fermionic Fock space spanned by ket vectors of the form
\be
\ket{\mathbf{\tau}} = \prod_{i=1}^L \left(c_i^\dagger\right)^{\tau_i} \ket{\mathbf{0}},
\ee 
where the product is ordered such that $\cd_i$ with higher $i$ act first. The label $\mathbf{\tau}=\{\tau_1,\ldots,\tau_L\}$, with $\tau_i=1$ if there is a fermion at site $i$ and $\tau_i=0$ if there is a hole. The vacuum state is defined as usual $c_i  \ket{\mathbf{0}} =0$ for $i=1,\ldots, L$.
%
%
Both models are supersymmetric chain models where the nilpotent supercharges $Q$, $Q^\dagger$ are built as sums of local operators. 

Originally, the FS model was considered with open boundary conditions \cite{FS07} and the FGNR model with periodic boundary conditions \cite{GFNR15}. In this section we give a short overview of \cite{FS07} and \cite{GFNR15}.
In Section \ref{sec:mapping} we restrict ourselves to the open boundary conditions for both models with even $L$ and discuss the mapping between them.

\subsection{FS model definition}
\label{sec:FSModelDef}
\begin{figure}
\begin{center}
\begin{tikzpicture}[scale=0.7,baseline=(current bounding box.center)]
\foreach\x in {0,...,8}{
\draw[thick] (\x,0) -- (\x+0.5,1);
}
\foreach\x in {1,...,8}{
\draw[thick] (\x-0.5,1) -- (\x,0);
}
\foreach \x in {0,...,8} {
\filldraw[fill=white, draw=black] (\x,0) circle (0.1);  
\filldraw[fill=white, draw=black] (\x+1/2,1) circle (0.1);  
}
\filldraw[fill=black, draw=black] (1,0) circle (0.1);  
\filldraw[fill=black, draw=black] (4,0) circle (0.1);  
\filldraw[fill=black, draw=black] (7,0) circle (0.1);  
\filldraw[fill=black, draw=black] (1+1/2,1) circle (0.1);  
\filldraw[fill=black, draw=black] (6+1/2,1) circle (0.1);  
\draw [<-] (1/2,1+0.3) arc (180:0:0.47cm);
\node[above] at (1,1.7) {\small $-\ii$};
\draw [<-] (1/2+2,1+0.3) arc (0:180:0.47cm);
\node[above] at (1+1,1.7) {\small $1$};
\draw [<-] (1/2+5,1+0.3) arc (180:0:0.47cm);
\node[above] at (1+5,1.7) {\small $1$};
\node[above] at (1+1+5,1.7) {\small $\ii$};
\draw [<-] (1/2+5+2,1+0.3) arc (0:180:0.47cm);
\node[above] at (1/2,-1.4) {\small $1$};
\draw [<-] (0,0-0.3) arc (-180:0:0.47cm);
\node[above] at (1+1/2,-1.4) {\small $\ii$};
\draw [<-] (2,0-0.3) arc (0:-180:0.47cm);
\draw [<-] (0+3,0-0.3) arc (-180:0:0.47cm);
\node[above] at (1/2+3,-1.4) {\small $1$};
\draw [<-] (2+3,0-0.3) arc (0:-180:0.47cm);
\node[above] at (1+1/2+3,-1.4) {\small $1$};
\draw [<-] (0+6,0-0.3) arc (-180:0:0.47cm);
\node[above] at (1/2+6,-1.4) {\small $-\ii$};
\draw [<-] (2+6,0-0.3) arc (0:-180:0.47cm);
\node[above] at (1+1/2+6,-1.4) {\small $1$};
\end{tikzpicture}
\caption{Statistical interaction between the two chains: Filled dots represent fermions, empty dots empty sites. The hopping amplitudes depend on the occupation of the other chain.}
	\label{fig:ActionOfH}
	\end{center}
\end{figure}
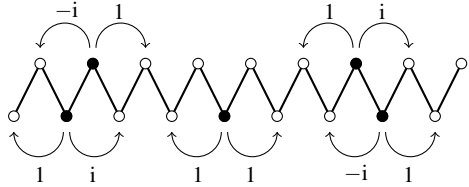

In this section we give a short overview of the FS model \cite{FS07}.
Consider the following supersymmetry generator
\begin{equation}
\label{eq:QFS1}
Q_{FS} = c_2^\dagger c_1 + \sum_{k=1}^{L/2-1} \left( e^{\ii \tfrac{\pi}{2} \alpha_{2k-2}}c^\dagger_{2k-1} + e^{\ii \tfrac{\pi}{2} \alpha_{2k}}c^\dagger_{2k+1} \right) c_{2k},
\end{equation}
where
\begin{equation}
\alpha_k = \sum_{j=1}^k (-1)^j n_j.
\end{equation}
This supersymmetry generator is nilpotent
\begin{equation}
Q_{FS}^2 = \left( Q_{FS}^\dagger \right)^2 = 0.
\end{equation}
The Hamiltonian is built up in the usual way
\begin{align}
H_{FS} &= \{ Q_{FS} ,\, Q_{FS}^\dagger \}
\nonumber \\ 
 & = \sum_{j=1}^{L-1} (c^\dagger_{j+1}p_j c_{j-1} + c^\dagger_{j-1}p_j c_{j+1} + \ii c^\dagger_{j+1}n_j c_{j-1} - \ii c^\dagger_{j-1}n_j c_{j-1})
 \nonumber \\
 & \ \ \ \ -2\sum_{j=1}^{L-1} n_j n_{j+1} + 2 F_1 + 2 F_2 + H_{bndry},
\end{align}
where
\begin{align}
 F_1 =\sum_{j=1}^{L/2} n_{2j-1}, \quad F_2 =\sum_{j=1}^{L/2} n_{2j}, \quad H_{bndry} = -n_1 - n_L.
\end{align}
This model describes an effective one-dimensional model where the fermions are hopping on two chains. Conveniently, these chains are denoted with odd and even site indices and they are coupled in a zig-zag fashion. The $F_1$ and $F_2$ operators are counting the fermions on the two chains, and $H_{FS}$ is block diagonal in these operators. The interaction between the chains is statistical: the hopping amplitude picks up an extra $\ii$ or $-\ii$ factor, if the fermion "hops over" another fermion on the other chain, see Fig. \ref{fig:ActionOfH}. There is a further attractive interaction between the two chains. 

The model is defined on an open chain, where the boundary interaction is encoded in $H_{bndry}$. 
The model can be shown to be solvable by nested coordinate Bethe Ansatz \cite{FS07}. The spectrum of the model is of the same form as for the free model which is defined by
\begin{equation}
H_{free} = 2 F_1 + 2F_2 + \sum_{j=2}^{L-1}\left( c_{j+1}^\dagger c_{j-1} + c_{j-1}^\dagger c_{j+1} \right) + H_{bndry}.
\end{equation}
The eigenenergies of $H_{FS}$ and $H_{free}$ are
\begin{align}
E &=2 \sum_{a=1}^{f_1+f_2} \left(1+\cos (2 p_a) \right),
\label{eq:E} \\
p_a &= m_a \frac{\pi}{L+1}, \qquad m_a \in \{1, 2, \ldots, L/2 \},
\label{eq:QuantCond}
\end{align}
where $p_a$ are called momenta, $f_1$ and $f_2$ are the number of fermions on the respective chains, i.e. the eigenvalues of $F_1$ and $F_2$.

The difference between the free and the interacting model is the \emph{degeneracy} of the energy levels. For the free model the Pauli principle is realized by fermions on the same chain not sharing the same momentum. The same momentum can be shared by two fermions on the two chains.  Hence for an eigenenergy characterized by the set $\{ m_a \}_{a=1}^{f_1+f_2}$ there are $\binom{L/2}{f_1}\binom{L/2}{f_2}$ possible choices, giving the degeneracy for the free model. 
For the interacting chain instead of thinking in terms of fermions it worth to consider \emph{exclusons} and \emph{Cooper pairs}. Exclusons are fermionic excitations satisfying quantization condition  (\ref{eq:QuantCond}) with the further restriction that an excluson prohibits any other particle to have the same momentum  $p_a$. 
A pair of fermions located on different chains can form a Cooper pair. In this case, two of the momenta (say $p_1$ and $p_2$) do not satisfy quantization condition (\ref{eq:QuantCond}) and instead they obey
\begin{equation}
\cos^2 p_1 = -\cos^2 p_2.
\end{equation}
The net energy contribution of the Cooper pair to (\ref{eq:E}) is zero. 

The spectrum of the FS model is built up as follows: there are $f_1$ and $f_2$ fermions on the respective chains. Out of these fermions $2C$ form Cooper pairs and $N_1 = f_1-C$, $N_2 = f_2-C$ are exclusons on the respective chains. An energy level  $E$ is characterized by the quantum numbers  $\{ m_a \}_{a=1}^{N_1+N_2}$ has the degeneracy
\begin{equation}
d_E = \binom{N_1+N_2}{N_1}\binom{L/2-N_1 - N_2}{C}.
\label{eq:IntDeg}
\end{equation}
The first term counts the possible distributions of the exclusons (with fixed $\{ m_a \}_{a=1}^{N_1+N_2}$ quantum numbers) on the two chains. The second term counts the degeneracy of the Cooper pairs. The interpretation of the second piece is that the Cooper pairs can be thought of indistinguishable quasiparticles like the exclusons and there is one possible Cooper pair for each allowed momentum. Moreover, the presence of an excluson with a given momentum prohibits a Cooper pair from occupying the corresponding level. This gives the spectrum and the degeneracy of the FS model. For further details we suggest the original publication \cite{FS07}. 
\subsection{FGNR model definition}
\label{sec:FGNRModelDef}
In this section we define the FGNR model \cite{GFNR15}.
%
%
%
%
We consider a one-dimensional supersymmetric lattice model which is a fermion-hole symmetric extension of the $M_1$ model of \cite{FSB03}. For this purpose define the operators $d_i^\dagger$ and $e_i$ by
\be
d_i^\dagger = p_{i-1}c_i^\dagger p_{i+1},\qquad e_i =  n_{i-1}c_i n_{i+1}. 
\ee
Hence $d^\dagger_i$ creates a fermion at position $i$ provided all three of positions $i-1$, $i$ and $i+1$ are empty. Similarly, $e_i$ annihilates a fermion at position $i$ provided $i$ and its neighbouring sites are occupied, i.e.
\begin{align}
d^\dagger_i \ket{\tau_1\ldots \tau_{i-2}\, 000\,\tau_{i+2}\ldots\tau_L} &= (-1)^{\mathcal{N}_{i-1}}\ket{\tau_1\ldots\tau_{i-2}\, 010\,\tau_{i+2}\ldots\tau_L},
\nonumber \\
e_i \ket{\tau_1\ldots\tau_{i-2}\, 111\,\tau_{i+2}\ldots\tau_L} &= (-1)^{\mathcal{N}_{i-1}}\ket{\tau_1\ldots\tau_{i-2}\, 101\,\tau_{i+2}\ldots\tau_L},
\end{align}
while these operators nullify all other states. Here $\mathcal{N}_i$ is the number operator. It counts the number of fermions to the left of site $i$. 
\begin{equation}
\mathcal{N}_i = \sum_{j=1}^i n_j,\qquad \mathcal{N}_{\rm F} = \mathcal{N}_L
\end{equation}
where $\mathcal{N}_{\rm F}$ is the total fermion number operator. 

We now define the nilpotent supersymmetric generators for the FGNR model
\be
Q_{FGNR}=\sum_{i=1}^L  (d_i^\dagger + e_i),\qquad Q_{FGNR}^2=0. 
\ee
The Hamiltonian is defined in the usual way
\begin{equation}
H_{FGNR}=\{Q_{FGNR}^\dagger,Q_{FGNR}\}.
\end{equation}
The Hamiltonian splits up naturally as a sum of three terms. The first term consists solely of $d$-type operators, the second solely of $e$-type operators and the third contains mixed terms.
\be
H_{FGNR}=H_I +H_{II}+H_{III},
\label{eq:FGNRmodeldef}
\ee
\begin{align}
H_I &= \sum_i \left( d_i^\dagger d_i + d_i d_i^\dagger\right) + \sum_i \left( d_{i}^\dagger d_{i+1} + d_{i+1}^\dagger d_i\right)
\nonumber\\
H_{II} &= \sum_i  \left( e_i e_i^\dagger + e_i^\dagger e_i\right)  +  \sum_i \left(e_i e_{i+1}^\dagger + e_{i+1} e_{i}^\dagger \right)
\nonumber \\
H_{III} &= \sum_i \left( e_i^\dagger d_{i+1}^\dagger\ + d_{i+1} e_i + e_{i+1}^\dagger d_i^\dagger + d_i e_{i+1}\right),
\label{eq:HII}
\end{align}
where we use periodic boundary conditions
\begin{equation}
c_{i+L}^\dagger = c_{i}^\dagger.
\end{equation}
Because the $d_i$'s and $e_i$'s are not simple fermion operators, they do not satisfy the canonical anticommutation relations.  As a result this bilinear Hamiltonian can not be diagonalized by taking linear combinations of $d$, $e$, $d^\dagger$ and $e^\dagger$.

The term $H_I$ alone is the Hamiltonian of the M$_1$ model of \cite{FSB03}. The addition of the operator $e_i$ introduces an obvious fermion-hole symmetry $d_i^\dagger \leftrightarrow e_i$ to the model.
It turns out that  this symmetry results in a surprisingly large degeneracy across the full spectrum of $H_{FGNR}$.

Note that the Hamiltonians $H_I$ and $H_{II}$ each contain only number operators and hopping terms and thus conserve the total number of fermions. The third Hamiltonian $H_{III}$ breaks this conservation law. For example, the term $e_i^\dagger d_{i+1}^\dagger$ sends the state $\ket{\ldots 1000\ldots}$ to  $\ket{\ldots 1110\ldots}$, thus creating two  fermions. Hence the fermion number is not conserved and therefore is not a good quantum number. However, the number of interfaces or domain walls between fermions and holes is conserved and we shall therefore describe our states in terms of these.
\subsubsection{Domain walls}
\label{sec:domainwalls}
We call an interface between a string of 0's followed by a string of 1's a 01-domain wall and a string of 1's followed by a string of 0's, a 10-domain wall. For example, assuming periodic boundary conditions the configuration
\[
000\Big| 11\Big| 000\Big| 1\Big| 0000\Big| 111\Big| , 
\]
contains six domain walls, three of each type and starting with a 01-domain wall.
Let us consider the effect of various terms appearing in (\ref{eq:HII}). As already discussed in an example above, the terms in $H_{III}$ correspond to hopping of domain walls and map between the following states
\be
\ket{\ldots 1\Big|000\ldots} \leftrightarrow  \ket{\ldots 111\Big|0\ldots},\qquad  \ket{\ldots 0\Big|111\ldots} \leftrightarrow  -\ket{\ldots 000\Big|1\ldots},
\label{eq:process1}
\ee
where the minus sign in the second case arises because of the fermionic nature of the model. Hopping of a domain wall always takes place in steps of two hence the parity of the positions of the domain walls is conserved. Aside from their diagonal terms, $H_I$ and $H_{II}$ correspond to hopping of single fermions or holes and therefore to hopping of \textit{pairs} of domain walls. They give rise to transitions between the states
\be
\ket{\ldots 0\Big|1\Big|00\ldots} \leftrightarrow  \ket{\ldots 00\Big|1\Big|0\ldots},\qquad  \ket{\ldots 1\Big|0\Big|11\ldots} \leftrightarrow  -\ket{\ldots 11\Big|0\Big|1\ldots}.
\label{eq:oddprocess}
\ee
Note that in these processes the total parity of positions of interfaces is again conserved, i.e. all processes in $H_{FGNR}$ conserve the number of domain walls at even and odd positions separately.

Finally, the diagonal term $\sum_i (d_i^\dagger d_i + d_i d_i^\dagger + e_i^\dagger e_i + e_i e_i^\dagger)$ in $H_{I}$ and $H_{II}$ counts the number of  $010$, $000$, $111$ and $101$ configurations. In other words they count the number of pairs of second neighbour sites that are both empty or both occupied
\be
\sum_i (d_i^\dagger d_i + d_i d_i^\dagger + e_i^\dagger e_i + e_i e_i^\dagger) = \\\sum_i (p_{i-1}p_{i+1} + n_{i-1}n_{i+1}).
\ee
This is equivalent to counting the total number of sites minus twice the number of domain walls that do not separate a single fermion or hole, i.e. twice the number of well separated domain walls. 

Since the number of odd and even domain walls is conserved the Hilbert space naturally breaks into sectors labeled by $(m, k)$, where $m$ is the total number of domain walls, and $k$ the number of odd domain walls. Due to periodic boundary conditions $m$ is even. 

\subsubsection{Solution of the FGNR model by Bethe ansatz}
\label{sec:BAforFGNR}
The FGNR model with periodic boundary conditions is solved by coordinate Bethe Ansatz \cite{GFNR15}. There are two kinds of conserved particles (even and odd domain walls), and hence the model is solved by a nested version of the Bethe Ansatz. A solution in the $(m, k)$ sector (with $m-k$ even and $k$ odd domain walls) is characterized by the Bethe roots $\{z_1, \ldots, z_m; u_1, \ldots, u_k \}$. In other words, $z$-type Bethe roots are associated with both kinds of domain walls, and $u$ type of Bethe roots are associated with odd domain walls. The complex numbers $\{z_1, \ldots, z_m; u_1, \ldots, u_k \}$ satisfy the following Bethe equations,
  \begin{align}
    z_j^L & = \pm\ii^{-L/2} \prod_{l=1}^k \frac{u_l-(z_j-1/z_j)^2}{u_l+(z_j-1/z_j)^2},\qquad j=1,\ldots,m \quad (m \in 2\mathbb{N}) ,
		\label{eq:BetheEq1} \\
    1 &=  \prod_{j=1}^{m} \frac{u_l-(z_j-1/z_j)^2}{u_l+(z_j-1/z_j)^2},\qquad l=1,\ldots,k,
		\label{eq:BetheEq2}
  \end{align}
where the $\pm$ is the same for all $j$.
Solutions corresponding to a nonzero Bethe vector are such that
\begin{align}
z_i^2 &\neq z_j^2, \quad \forall i,j \in \{1,\ldots, m \}, 
\nonumber \\
u_i &\neq u_j, \quad \forall i,j \in \{1,\ldots, k \}.
\end{align}
Two solutions $\{z_1, \ldots, z_m; u_1, \ldots, u_k \}$, $\{z'_1, \ldots, z'_m; u'_1, \ldots, u'_k \}$ lead to the same Bethe vector if there exist two permutations $\pi \in S_m$ and $\sigma \in S_k$ and a set of $\mathbb{Z}_2$ numbers $\{\epsilon_i\}_{i=1}^m$ (i.e. $\epsilon_j=\pm 1$) such that $z_j=\epsilon_j z'_{\pi(j)}$ and $u_l = u'_{\sigma(l)}$. 
The eigenenergy corresponding to the Bethe roots $\{z_1, \ldots, z_m; u_1, \ldots, u_k \}$ is
\begin{equation}
 \Lambda =L+ \sum_{i=1}^{m} (z_i^2+z_i^{-2}-2),
\end{equation}
which in fact depends only on the non-nested $z$ Bethe roots. 
The Bethe equations have free fermionic solutions in the following cases. When $k=0$ there are  no Bethe equations at the nested level and the first set of equations simplifies to the free fermionic case. When $k=1$ the $u=0$ solutions give the free fermionic part of the spectrum. It is worth to note that the spectrum of the FGNR model with periodic boundary conditions does have a non free fermionic part. This part will not transfer to the open boundary case. 

\subsubsection{Cooper pairs}
Consider a free fermionic solution
\begin{equation}
z_j = \ii^{-1/2} \e^{2\ii \pi I_j / L}, \quad j=1,\ldots, m,
\label{eq:FFsol}
\end{equation}
where $I_j$ is a (half-)integer. This solves the Bethe equations for the $k=0$ case, or the $k=1$ case with $u=0$. This same solution can be used to find a solution in the sector with two more odd domain walls (with $k=2$ and $3$ in the respective cases). Consider the $k=2$ case. Bethe equations (\ref{eq:BetheEq1}) with $k=2$ are solved by (\ref{eq:FFsol}) if the nested Bethe roots $u_1$, $u_2$ satisfy  
\begin{equation}
u_2 = -u_1
\end{equation}
as in this case the two new terms in the first Bethe equations cancel each other
\begin{multline}
\frac{u_1-(z_j-1/z_j)^2}{u_1+(z_j-1/z_j)^2}\cdot\frac{u_2-(z_j-1/z_j)^2}{u_2+(z_j-1/z_j)^2}  \\ =\frac{u_1-(z_j-1/z_j)^2}{u_1+(z_j-1/z_j)^2}\cdot\frac{u_1+(z_j-1/z_j)^2}{u_1-(z_j-1/z_j)^2}=1.
\end{multline}
Hence the first Bethe equations (\ref{eq:BetheEq1}) with solution (\ref{eq:FFsol}) serve as a consistency condition for $u_1,\, u_2$
	\begin{equation}
		  1 =  \prod_{j=1}^{m} \frac{u_l-(z_j-1/z_j)^2}{u_l+(z_j-1/z_j)^2},\qquad l=1,2.
		\end{equation}
For a free fermionic solution there are always (purely imaginary) $u_2 = -u_1$ type nested roots. As this solution has the same $z$ type roots a solution with the Cooper pair has the same  energy $\Lambda$ as the original one.
We can continue like this introducing new Cooper pairs. The creation of Cooper pairs is limited by the number of domain walls. 
For further details we suggest \cite{GFNR15}.

\section{Mapping between the domain wall and particle representations}
\label{sec:mapping}
The definition of the open boundary version of the FGNR model is straightforward. 
In the open boundary version with a system size $L$ the operators $d_i, d_i^\dagger$ and $e_i^\dagger, e_i$ are defined for $i = 1, \ldots, L-1$. When $i=1$ we need to introduce the extra $0$-th site and fix its state to be empty. With these definitions the open boundary supercharge
\begin{equation}
Q_{FGNR}^{(OBC)} = \sum_{i=1}^{L-1}\left( d_i^\dagger + e_i \right)
\end{equation}
is well defined and nilpotent. In this section we would like to lighten the notation for the supercharges and work with their conjugated counterparts. We also need to introduce an intermediate model given in terms of the operators
\begin{align*}
&g_i = p_{i+1} c_i n_{i-1}, \qquad f_i = n_{i+1} c_i p_{i-1}.
\end{align*}

Fix $L$ to be even. We have the following supercharges
\begin{align}
&\Qd_L = (Q_{FGNR}^{(OBC)})^{\dag} =\sum_{i=1}^{L-1} \left( d_{i}+e_{i}^{\dag}\right), \label{QFGNR} \\
&\Qtd_L = Q_{FS}^{\dag} = \cd_1 c_2 + \sum_{i=1}^{L/2-1}\cd_{2i+1} \left( c_{2i} e^{{\rm i} \alpha_{2i-1}\pi/2} + c_{2i+2} e^{{\rm i} \alpha_{2i}\pi/2}\right)  \label{QFS},
\end{align}
We also define two additional supercharges $\Qd_{e,L}$ and $\Qd_{o,L}$ which represent hopping of domain walls and are required as an intermediate step of the mapping,
\begin{align} 
&\Qd_{e,L} = \sum_{i=1}^{L/2-1} \left( g_{2i}+f_{2i}^{\dag}\right), \label{Qeven} \\
&\Qd_{o,L} =  g_{1}+f_{1}^{\dag}+\sum_{i=1}^{L/2-1} \left( g_{2i+1}+f_{2i+1}^{\dag}\right), \label{Qodd}
\end{align}
Notice that the first terms in the summations in the charges $\Qd_{L}$ and $\Qd_{o,L}$ contain $n_0$. As mentioned above, we need to fix the $0$-th site to be unoccupied. Hence the eigenvalue of $n_0$ is always $0$. 

We would like to find the map between $\Qd_{L}$ and $\Qtd_L$. Assume that it is given by a transformation $T$,
\begin{align}\label{TQT}
&\Qtd_L = T\Qd_{L} T^{\dag}, \qquad T= P \G M, 
\end{align}
which itself consists of three terms. The first operator $M$ turns creation and annihilation of domain walls into hopping of domain walls. The second operator $\G$ turns domain walls into particles and the third operator $P$ fixes the phase factors such that they match (\ref{QFS}). 
 
Now we turn to the discussion of the transformations $M$, $\G$ and $P$. 

\subsection{Transformation $M$}
\label{se:Mtransform}
The first term $M$ translates to the dynamics of domain walls, i.e. it transforms linear combinations of the operators $d,d^{\dag}$ and $e,e^{\dag}$ into linear combinations of $f,f^{\dag}$ and $g,g^{\dag}$ (see \cite{GFNR15} for more details). More precisely:
\begin{align}\label{M}
M=\prod_{i=0}^{\lfloor (L-1)/4 \rfloor} (c_{4i+1}-\cd_{4i+1})(c_{4i+2}-\cd_{4i+2}),
\end{align}
and for all even $i$ we have
\begin{align}
&M(d_i+e^{\dag}_i) M^{\dag} = f_i +g^{\dag}_i,\qquad M(d_{i+1}+e^{\dag}_{i+1}) M^{\dag} = f^{\dag}_{i+1} +g_{i+1}.
\end{align}
In other words $M$ turns $\Qd$ into a combination of $Q_e$ and $\Qd_o$
\begin{align}
M \Qd_L M^{\dag} = Q_{e,L}+\Qd_{o,L}. \label{MQMeven}
\end{align}
Thus we get an intermediate model.


\subsection{Transformation $\G$}
\label{se:Gtransform}
The next transformation $\G$ turns domain walls into particles. In fact, there is a family of such transformations. We select $\G$ to be of the following form
\begin{align}
\G_L=\prod_{i=1}^{L-1}\left(p_i +n_i ( \cd_{i+1}+c_{i+1})\right). \label{GammaF}
\end{align}
This operator satisfies $\G \G^{\dag}=1$ and transforms the monomials in $c_i$ and $\cd_i$ of (\ref{MQMeven}) into those of $\Qtd$ (\ref{QFS}). More precisely, conjugation by $\G$ has the following effect
\begin{align}
\G_{L} ( Q_{e,L}+\Qd_{o,L}) \Gd_{L} = \Qhd_L, \label{eq:GQG}
\end{align}
where the new supercharge $ \Qhd_L$ differs from $\Qtd$ only by phase factors. 
Let us take $\G$ and act termwise on $Q_{e}+\Qd_{o}$, i.e. on the combination
\begin{align*}
f_i +g^{\dag}_i+f^{\dag}_{i+1} +g_{i+1},
\end{align*}
for the labels $i>0$ and separately on the first term $f^{\dag}_{1} +g_{1}$. We find
\begin{align}
&\G_{L}\left(f^{\dag}_{1} +g_{1}\right) \Gd_{L}= -\cd_1 c_2 ~ e^{{\rm i}\pi \sum_{j=1}^{L/2-1} n_{2j+1} }
 , \label{eq:Gamfg0}
\end{align}
and
\begin{multline}
\G_{L}\left(f_i +g^{\dag}_i+f^{\dag}_{i+1} +g_{i+1}\right) \Gd_{L} \\ =  
\left(\cd_{2i+1} c_{2i} 
~ e^{{\rm i}\pi \sum_{j=1}^{2i-1} n_{j} } - 
\cd_{2i+1} c_{2i+2} \right)
 e^{{\rm i}\pi \sum_{j=i+1}^{L/2-1} n_{2j+1} }.
\label{eq:Gamfgi}
\end{multline}
Therefore $\Qhd_L$ as defined in (\ref{eq:GQG}) becomes
\begin{multline}\label{eq:Qhat}
\Qhd_L =
-\cd_1 c_2 ~ e^{{\rm i}\pi \sum_{j=1}^{L/2-1} n_{2j+1} } \\ +
\sum_{i=1}^{L/2-1}
\left(\cd_{2i+1} c_{2i} 
e^{{\rm i}\pi \sum_{j=1}^{2i-1} n_{j} } - 
\cd_{2i+1} c_{2i+2} \right)
 e^{{\rm i}\pi \sum_{j=i+1}^{L/2-1} n_{2j+1} }.
\end{multline}
This agrees with  (\ref{QFS}) up to the phase factors.


\subsection{Transformation $P$}
\label{se:Ptransform}
Let $p(\nu_1,\nu_2,\dots,\nu_{L})$ be an unknown function of a binary string, $\nu_i=0,1$. Write a generic phase factor transformation $P_L$
\begin{align}
P_L=e^{{\rm i}\frac{\pi}{2} \hat{p}(n_1,n_2,\dots,n_{L})}, \label{eq:Pans}
\end{align}
and the function $p$ introduced above denotes the eigenvalue of $\hat{p}$ on the state $\ket{\nu_1,\dots,\nu_{L}}$.
The commutation relations
\begin{align}
\hat{p}(n_1,\dots,n_i,\dots,n_{L}) c_i = c_i \hat{p}(n_1,\dots,n_i-1,\dots,n_{L}), \\
\hat{p}(n_1,\dots,n_i,\dots,n_{L}) \cd_i = \cd_i \hat{p}(n_1,\dots,n_i+1,\dots,n_{L}),
\end{align}
hold on all states of the Hilbert space. Let us find $P$ such that
\begin{align}
\Qtd_L = P_{L} \Qhd_L P_{L}^{-1}.
\end{align}
Commuting $P_L$ through each monomial of $\Qhd_L$ and comparing it with the corresponding monomial of $\Qtd_L$ 
 we find $L-1$ conditions. Commuting $P_L$ with the first monomial in (\ref{eq:Qhat}) and acting on the state $\ket{\nu_1,\dots,\nu_{L}}$ leads to the first condition
\begin{align}\label{eq:Pi=0}
p(\nu_1+1,\nu_2-1,\dots,\nu_{L}) - p(\nu_1,\nu_2,\dots,\nu_{L})+ 2\sum_{j=1}^{L/2-1} \nu_{2j+1} +2 =0.
\end{align}
Commuting $P_L$ with the two bulk terms in (\ref{eq:Qhat}) leads to
\begin{multline}
\label{eq:Peqev}
p(\nu_1,\dots,\nu_{2i}-1,\nu_{2i+1}+1,\dots,\nu_{L}) - p(\nu_1,\dots,\nu_{L}) \\ + 2\sum_{j=1}^{2i-1} \nu_{j} + 2\sum_{j=i+1}^{L/2-1} \nu_{2j+1}  =  \sum_{j=1}^{2i-1} (-1)^j \nu_j , 
\end{multline}
and
\begin{multline}
\label{eq:Peqodd}
p(\nu_1,\dots,\nu_{2i+1}+1,\nu_{2i+2}-1,\dots,\nu_{L}) - p(\nu_1,\dots,\nu_{L}) \\
+ 2\sum_{j=i+1}^{L/2-1} \nu_{2j+1} +2 =  \sum_{j=1}^{2i} (-1)^j \nu_j.
\end{multline}
The second equation here with $i=0$ reproduces (\ref{eq:Pi=0}). In the second equation we can replace $\nu_{2i+1}$ with $\nu_{2i+1}-1$ and $\nu_{2i+2}$ with $\nu_{2i+2}+1$. As a result it becomes of the same form as the first one. 
Hence (\ref{eq:Peqev}) and (\ref{eq:Peqodd}) together define $L$ equations for $k=0,\dots,L-1$, where for even $k$ one uses (\ref{eq:Peqodd}) and for odd $k$ one uses (\ref{eq:Peqev}). These equations are valid modulo $4$ and can be further simplified
\begin{multline}
p(\nu_1,\dots,\nu_{2i}-1,\nu_{2i+1}+1,\dots,\nu_{L}) - p(\nu_1,\dots,\nu_{L})  \\
=  \sum_{j=1}^{2i-1} (-1)^{j+1} \nu_j + 2\sum_{j=i+1}^{L/2-1} \nu_{2j+1}  ,
\end{multline}
and
\begin{multline}
p(\nu_1,\dots,\nu_{2i+1}-1,\nu_{2i+2}+1,\dots,\nu_{L})-p(\nu_1,\dots,\nu_{L}) \\
=\sum_{j=1}^{2i} (-1)^{j+1} \nu_j+ 2\sum_{j=i+1}^{L/2-1} \nu_{2j+1} +2.
\end{multline}
These two equations can be united into one equation using one index $k$ which can be odd or even 
\begin{multline}
p(\nu_1,\dots,\nu_k,\nu_{k+1},\dots,\nu_{L})- p(\nu_1,\dots,\nu_k+1,\nu_{k+1}-1,\dots,\nu_{L})  \\
= w_k(\nu_1,\dots,\nu_{2N}), \label{eq:prec}
\end{multline}
where the right hand side is given by 
\begin{align}
&w_k(\nu_1,\dots,\nu_{L}) =(1-(-1)^k)+  \sum_{j=1}^{k-1} (-1)^{j+1} \nu_j + \sum_{j=k+2}^{L-1} (1-(-1)^j) \nu_{j} .\label{eq:wk}
\end{align}
Therefore we find a set of recurrence relations for the functions $p$. Note that for a given configuration with the binary string $(\nu_1,\dots,\nu_{L})$ these equations are assumed to hold for those values of $k$ for which $\nu_k=0$ and $\nu_{k+1}=1$. Hence the number of such equations is equal to the number of domain   walls of type $01$. 

\subsection{Particle position coordinates}
It is more natural to solve the equations \eqref{eq:prec} in a basis where the vectors are labelled using particle positions $x_k$.
The Hilbert spaces in both models are given by vectors labelled by strings of $L$ numbers $\tau_i=0,1$
\be
\ket{\tau_1,\ldots,\tau_{L}}_n =\ket{\mathbf{\tau}} = \prod_{i=1}^{L} \left(c_i^\dagger\right)^{\tau_i} \ket{\mathbf{0}}.
\ee 
We attached the subscript $n$ in the above notation in order to distinguish it from another labelling of the vectors in the same Hilbert space. Let $m$ be the number of particles in the system. Let us introduce a basis labelled by the positions of the particles and let $\rho$ denote the mapping between the two labellings 
\begin{align}
\label{rhodef}
\rho: \ket{\nu_1,\dots,\nu_{L}}_n \mapsto \ket{x_1,\dots,x_m}_x.
\end{align}
The numbers $x_k$ are the eigenvalues of the operators $\hat{x}_k$ which coincide with the eigenvalues of the operators $j n_{j}$ with $j=x_k$. 

Fix $m$ to be the total number of particles in the system and define two functions $\tilde{p}$ and $\tilde{w}$ using the mapping $\rho$
\begin{align*}
&\rho\left( \hat{p}(n_1,\dots,n_{L})\ket{\nu_1,\dots,\nu_{L}}_n \right) = \tilde{p}(x_1,\dots,x_m) \ket{x_1,\dots,x_m}_x, \\
&\rho\left( \hat{w}_k(n_1,\dots,n_{L})\ket{\nu_1,\dots,\nu_{L}}_n \right)= \tilde{w}_k(x_1,\dots,x_m) \ket{x_1,\dots,x_m}_x,
\end{align*}
with $ \hat{w}_k$ being the diagonal operator with the eigenvalues (\ref{eq:wk}).
We can now rewrite (\ref{eq:prec}) and (\ref{eq:wk}) in the particle position basis 
\begin{align}
\tilde{p}(x_1,\dots,x_j,\dots,x_m) - \tilde{p}(x_1,\dots,x_j-1,\dots,x_m) = \tilde{w}_j(x_1,\dots,x_{m}),
\label{eq:ptrec}
\end{align}
with
\begin{align}
\tilde{w}_k(x_1,\dots,x_{m}) = (1+(-1)^{x_k}) - \sum_{i=1}^{k-1} (-1)^{x_i} + \sum_{i=k+1}^{m} (1- (-1)^{x_i}) .
\label{eq:wtk}
\end{align}
Once again this equation is considered to hold for $j$ such that $x_{j}-x_{j-1}>1$. The generic solution of (\ref{eq:ptrec}) is 
\begin{align}
& \tilde{p}(x_1, \ldots,x_m)  =\tilde{p}(1,2,\ldots,m)+
\sum_{k=1}^m \sum_{i=k+1}^{x_k} \tilde{w}_{k}(1,2,\ldots,k-1,i,x_{k+1},\ldots,x_m),
 \end{align}
which can be checked by a direct calculation. 
Here $\tilde{p}(1,2,\ldots,m)$ is the initial condition and can be chosen to be $0$. Inserting $\tilde{w}_k$ we get
\begin{align}
\tilde{p}(x_1,\ldots,x_m)  =
&\sum_{k=1}^m
 \sum_{i=k+1}^{x_k} \left( 1+(-1)^i -\sum_{j=1}^k (-1)^j + \sum_{j=k+1}^m (1-(-1)^{x_j}) \right).
\end{align}
The required phase transformation takes the form
\begin{align}
P_L=e^{{\rm i}\frac{\pi}{2} \left(\tilde{p}(\hat{x}_1,\dots,\hat{x}_m)\circ\rho \right)}. \label{eq:Psol}
\end{align}

\subsection{Examples}
\label{se:examples}

To illustrate the mapping, we show some examples of corresponding states between the FS model (zig-zag ladder) and FGNR model states (up to phase factors $\pm 1$, $\pm i$),
\begin{align}
& {\rm empty}\ {\rm ladder}: & |0000 \, 0000 \, 0000 \rangle_{FS} & \quad \leftrightarrow & {}_0 |1100 \, 1100 \, 1100 \rangle_{FGNR}
\nonumber \\
& {\rm single}\ {\rm FS\ semion}: & |0000 \,1000 \, 0000 \rangle_{FS} & \quad \leftrightarrow & {}_0 |1100 \, 0011 \, 0011 \rangle_{FGNR}
\nonumber \\
& {\rm single}\ {\rm FS\ pair}: & |0000 \,1100 \, 0000 \rangle_{FS} & \quad \leftrightarrow & {}_0 |1100 \, 0100 \, 1100 \rangle_{FGNR}
\nonumber \\
& {\rm lower}\ {\rm leg}\ {\rm filled}: & |1010 \, 1010 \, 1010 \rangle_{FS} & \quad \leftrightarrow & {}_0 |0000 \, 0000 \, 0000 \rangle_{FGNR}
\nonumber \\
& {\rm single}\ {\rm FGNR\ particle}: & |1010 \, 0110 \, 1010 \rangle_{FS} & \quad \leftrightarrow & {}_0 |0000 \, 1000 \, 0000 \rangle_{FGNR}
\nonumber \\
& {\rm upper}\ {\rm leg}\ {\rm filled}: & |0101 \, 0101 \, 0101 \rangle_{FS} & \quad \leftrightarrow & {}_0 |1010 \, 1010 \, 1010 \rangle_{FGNR}
\nonumber \\
& {\rm upper}\ {\rm leg}\ {\rm plus}\ {\rm semion}: & |1101 \, 0101 \, 0101 \rangle_{FS} & \quad \leftrightarrow & {}_0 |0101 \, 0101 \, 0101 \rangle_{FGNR}
\nonumber \\
& {\rm filled}\ {\rm ladder}: & |1111 \, 1111 \, 1111 \rangle_{FS} & \quad \leftrightarrow & {}_0 |0110 \, 0110 \, 0110 \rangle_{FGNR}
\nonumber
\end{align}

\noindent
For $L=4$ the phase factors take the explicit values 
\begin{align}
& \tilde{p}(1)=0, \ \  \tilde{p}(2)=2, \ \ \tilde{p}(3)=0, \ \ \tilde{p}(4)=0
\nonumber \\  
& \tilde{p}(1,2)=0, \ \  \tilde{p}(1,3)=1, \ \ \tilde{p}(1,4)=0, \ \ \tilde{p}(2,3)=1, \ \ \tilde{p}(2,4)=2, \ \ \tilde{p}(3,4)=2
\nonumber \\
& \tilde{p}(1,2,3)=0, \ \ \tilde{p}(1,2,4)=2, \ \ \tilde{p}(1,3,4)=3, \ \ \tilde{p}(2,3,,4)=3
\nonumber \\
& \tilde{p}(1,2,3,4) = 0
\end{align}

Finally we provide an explicit example of all the steps in the mapping. Let us act with both sides of (\ref{TQT}) on the state 
$\ket{010101}$ 
\begin{align*}
\Qtd_6 \ket{010101} =P \G M \Qd_{6} M^{\dag}\G^{\dag} P^{-1}\ket{010101},
\end{align*}
The action of $\Qtd_6$ results in
\begin{align*}
\Qtd_6 \ket{010101} =
&\ket{001101} +\ii \ket{010011} -\ket{010110} +\ii \ket{011001}
   +\ket{100101},
\end{align*}
and the right hand side is computed as follows
\begin{align*}
&P \G M \Qd_{6} M^{\dag}\G^{\dag} P^{-1}\ket{010101}=
-P \G M \Qd_{6} M^{\dag}\G^{\dag} \ket{010101} \\
&=P \G M \Qd_{6} M^{\dag}\ket{011001}=
P \G M \Qd_{6}\ket{101010}\\
&=
P \G M  \big{(}
\ket{001010} -\ket{100010} +\ket{101000} +\ket{101110} -\ket{111010} \big{)}
   \\
&= P \G  \big{(}
\ket{001001}  - \ket{010001} +\ket{011011} +\ket{011101}
   +\ket{111001}\big{)}\\
 &=P
 \big{(}
 \ket{001001} -\ket{010001} +\ket{011011} +\ket{011101}
   +\ket{111001}
\big{)}\\
&=
\ket{001101} +\ii \ket{010011} -\ket{010110}+\ii \ket{011001}
   +\ket{100101}.
\end{align*}

\section{Conclusion}
\label{sec:conclusion}

We have established a unitary transformation between the FS and FGNR models with open boundary conditions. We are confident that this map will be helpful for unraveling the properties of these highly intriguing models. For example, in the FS formulation it was not clear how to impose periodic boundary conditions without losing the supersymmetry - this issue is now resolved.

It is a pleasure to dedicate this work to Bernard Nienhuis on the occasion of his 65th birthday. Bernard has always had a special eye for ingenious maps relating apparently different models of statistical physics to one another. We can only hope that dedicating this work to him finds some justification in our following a similar strategy for one of the many integrable models that he has pioneered.

\section{Acknowledgment}
\label{sec:acknowledgment}
We thank the hospitality of the international mathematical research institute MATRIX where a large part of this work was performed. JdG and AG gratefully thank financial support of the ARC Centre of Excellence for Mathematical and Statistical Frontiers (ACEMS).

\newcommand{\arxiv}[1]{\href{http://arxiv.org/abs/#1}{arXiv/#1}}

\end{document}